\documentclass[twoside,journey]{IEEEtran}

\usepackage{makecell}
\usepackage{hyperref}
\usepackage{array}
\usepackage{caption2}
\usepackage{graphicx,amssymb,amsmath}
\usepackage{multicol}
\usepackage[noadjust]{cite}
\usepackage{setspace}
\usepackage{subfigure}
\usepackage{graphicx}
\usepackage{float}
\usepackage{url}
\usepackage{stfloats}
\usepackage{amsthm,pifont}
\usepackage{flushend}
\usepackage{cases,subeqnarray}
\usepackage{bm,multirow,bigstrut}
\usepackage{amsmath, amsthm, amssymb}
\usepackage{textcomp}
\usepackage{latexsym,bm}
\usepackage{booktabs}
\usepackage{xcolor}
\usepackage{mathtools}
\usepackage{dsfont}
\usepackage{extarrows}
\usepackage{epsfig}
\usepackage{epsfig}
\usepackage{epstopdf}
\usepackage[noend]{algpseudocode}
\usepackage{algorithmicx,algorithm}

\theoremstyle{plain}

\theoremstyle{plain}

\usepackage{amsmath}

\IEEEoverridecommandlockouts
\begin{document}
\title{Semantic Communications for Artificial Intelligence Generated Content (AIGC) Toward Effective Content Creation}

\author{Guangyuan Liu, Hongyang Du, Dusit Niyato,~\IEEEmembership{Fellow,~IEEE}, Jiawen~Kang, Zehui~Xiong, \\
Dong In Kim,~\IEEEmembership{Fellow,~IEEE}, and Xuemin~(Sherman)~Shen,~\IEEEmembership{Fellow,~IEEE}

\thanks{This research is supported by the National Research Foundation, Singapore, and Infocomm Media Development Authority under its Future Communications Research \& Development Programme, DSO National Laboratories under the AI Singapore Programme (AISG Award No: AISG2-RP-2020-019 and FCP-ASTAR-TG-2022-003), Energy Research Test-Bed and Industry Partnership Funding Initiative, Energy Grid (EG) 2.0 programme, DesCartes and the Campus for Research Excellence and Technological Enterprise (CREATE) programme, and MOE Tier 1 (RG87\/22).}
\thanks{The work is also supported by NSFC under grant No. 62102099, U22A2054, and the Pearl River Talent Recruitment Program under Grant 2021QN02S643, and Guangzhou Basic Research Program under Grant 2023A04J1699}
\thanks{The research is also supported by the National Research Foundation (NRF) and Infocomm Media Development Authority under the Future Communications Research Development Programme (FCP). The research is also supported by the SUTD SRG-ISTD-2021-165, and the Ministry of Education, Singapore, under its SMU-SUTD Joint Grant (22-SIS-SMU-048).}
\thanks{This research is also supported in part by the MSIT (Ministry of Science and ICT), Korea, under the ICT Creative Consilience program (IITP-2020-0-01821) supervised by the IITP (Institute for Information \& Communications Technology Planning \& Evaluation).}

\thanks{G.~Liu, and H.~Du are with the School of Computer Science and Engineering, the Energy Research Institute @ NTU, Interdisciplinary Graduate Program, Nanyang
Technological University, Singapore (e-mail: liug0022@e.ntu.edu.sg, hongyang001@e.ntu.edu.sg).}
\thanks{D. Niyato is with the School of Computer Science and Engineering, Nanyang Technological University, Singapore (e-mail: dniyato@ntu.edu.sg).}
\thanks{J. Kang is with the School of Automation, Guangdong University of Technology, China. (e-mail: kavinkang@gdut.edu.cn).}
\thanks{Z. Xiong is with the Pillar of Information Systems Technology and Design, Singapore University of Technology and Design, Singapore (e-mail:
zehui\_xiong@sutd.edu.sg).}
\thanks{D. I. Kim is with the Department of Electrical and Computer Engineering,
Sungkyunkwan University, South Korea (e-mail: dikim@skku.ac.kr).}
\thanks{X. Shen is with the Department of Electrical and Computer Engineering, University of Waterloo, Canada (e-mail: sshen@uwaterloo.ca).}
}
\maketitle
\vspace{-1cm}
\begin{abstract}
Artificial Intelligence Generated Content (AIGC) Services have significant potential in digital content creation. The distinctive abilities of AIGC, such as content generation based on minimal input, hold huge potential, especially when integrating with semantic communication (SemCom). In this paper, a novel comprehensive conceptual model for the integration of AIGC and SemCom is developed. Particularly, a content generation level is introduced on top of the semantic level that provides a clear outline of how AIGC and SemCom interact with each other to produce meaningful and effective content. Moreover, a novel framework that employs AIGC technology is proposed as an encoder and decoder for semantic information, considering the joint optimization of semantic extraction and evaluation metrics tailored to AIGC services. The framework can adapt to different types of content generated, the required quality and the semantic information utilized. By employing a Deep Q Network (DQN), a case study is presented that provides useful insights into the feasibility of the optimization problem and its convergence characteristics.
\end{abstract}

\section{Introduction}
Artificial Intelligence Generated Content services (AIGC services) have gained significant attention due to their ability to enhance creativity, accelerate design processes, provide personalized content and promote accessibility in various domains, such as digital marketing, video game design and filmmaking \cite{cao2023comprehensive}. AIGC employs advanced Generative AI (GAI) techniques to produce content corresponding to human instructions by deciphering the intent and generating appropriate content in response.

The considerable advancements in AIGC can be attributed to the development of increasingly sophisticated generative models, expanded foundation model architectures and the accessibility of vast computational resources. AIGC services hold substantial importance for a variety of reasons, including:

\begin{itemize}
\item \textbf{Enhancing creativity:} GAI models support users in generating unique and original images from textual prompts, thereby unveiling new avenues for artistic and creative expression\footnote{https://hbr.org/2022/11/how-generative-ai-is-changing-creative-work, accessed July 06, 2023}.
\item \textbf{Streamlining design processes:} AIGC services facilitate rapid generation of visual content, optimizing workflows for designers and artists which subsequently reducing time and effort required for manual design tasks.
\item \textbf{Tailoring content:} These services enable the creation of customized content that is specifically designed to cater to individual user preferences or distinct target audiences, ultimately resulting in heightened user engagement.
\item \textbf{Promoting accessibility:} AIGC services enable the users with limited design abilities to generate professional-quality visuals, thereby fostering inclusivity and democratizing the design process.
\end{itemize}
The efficient functioning and widespread adoption of AIGC services are contingent on robust communication infrastructure and technologies, such as wireless networks and semantic communication (SemCom). The technologies offer many benefits to AIGC services including~ \cite{du2023enabling}: 

\begin{itemize}
\item \textbf{Ubiquitous access:} Wireless networks facilitate widespread access to AIGC services, allowing users to benefit from these services regardless of their location or device.
\item \textbf{Real-time processing:} Wireless networks enable rapid data transfer, promoting real-time interactions between users and AI models for content generation, thus improving the overall user experience.
\item \textbf{Scalability:} Wireless networks can efficiently support a large number of users simultaneously, enhancing the availability and accessibility of AIGC services.
\end{itemize}

On the other hand, SemCom can focus on the intended meaning or context of the transmitted data, rather than on its raw form, which leads to a more efficient and intelligent exchange of information. When SemCom is applied to AIGC services, this results in optimized transmission and interpretation of generated content. Specifically, the benefits of integrating SemCom into AIGC services include \cite{lin2023unified}:
\begin{itemize}
\item \textbf{Bandwidth efficiency:} SemCom makes possible more efficient use of bandwidth, allowing generated contents to be transmitted with lower latency and less resource consumption.
\item \textbf{Enhanced privacy:} SemCom also contributes to increased privacy, ensuring secure exchanges of information between users and AIGC services due to shorter transmission time and less amount of transferred data.
\item \textbf{Energy Efficiency:} By focusing on the meaningful components of data, SemCom can reduce the energy consumption of data transmission and processing in AIGC services.
\end{itemize}
Evidently, the development of efficient and secure wireless and SemCom technologies is essential for the widespread adoption and effective implementation of AIGC services. As such, further clear understanding of their integration is necessary to guide the design and implementation of AIGC services to meet the diverse needs of different applications and domains.


In this paper, we introduce a conceptual model for integrating and designing AIGC services with SemCom. Specifically, we extend three-level of SemCom, i.e., physical level, semantic level and effectiveness level, by having a generation level that allows AIGC services at the receiver to generate meaningful contents from input provided by the transmitter over a communication link. To realize the proposed model, we then explore the effect of extracting semantic information from AIGC inputs on the AIGC output quality. We propose an innovative framework that quantitatively evaluates the impact of various factors, such as image resolution and compression levels. The contributions of this paper are summarized as follows:
\begin{itemize}
\item We present the model that serves as a foundation to develop AIGC services by leveraging the capability of SemCom. The model can represent building blocks of the SemCom-enabled AIGC services and their interactions to achieve effective and efficient content generation by generative AI over wireless networks. 
\item We realize the model by conducting a thorough analysis of how alterations in the extraction and reduction of input information, such as changes in image resolution and compression, influence AIGC output quality.
\item We introduce a novel framework that describes the vital role of AIGC in semantic information transmission. This strategy aims to optimize resource allocation to maximize user satisfaction with AIGC services.
\item We present an in-depth experimental study of joint resource allocation to understand the complexities of accommodating various services within a single transmission framework. This study offers crucial insights into optimization results and their convergence characteristics. Then, we provide validation of our proposed framework through a Deep Q-Network (DQN), demonstrating its feasibility and promising implications for enhancing AIGC services efficiency.
\end{itemize}

\section{Semantic Communications-Enabled AI-Generated Content Model}\label{sec2}
\begin{figure}[t!]
\centerline{\includegraphics[width=0.5\textwidth]{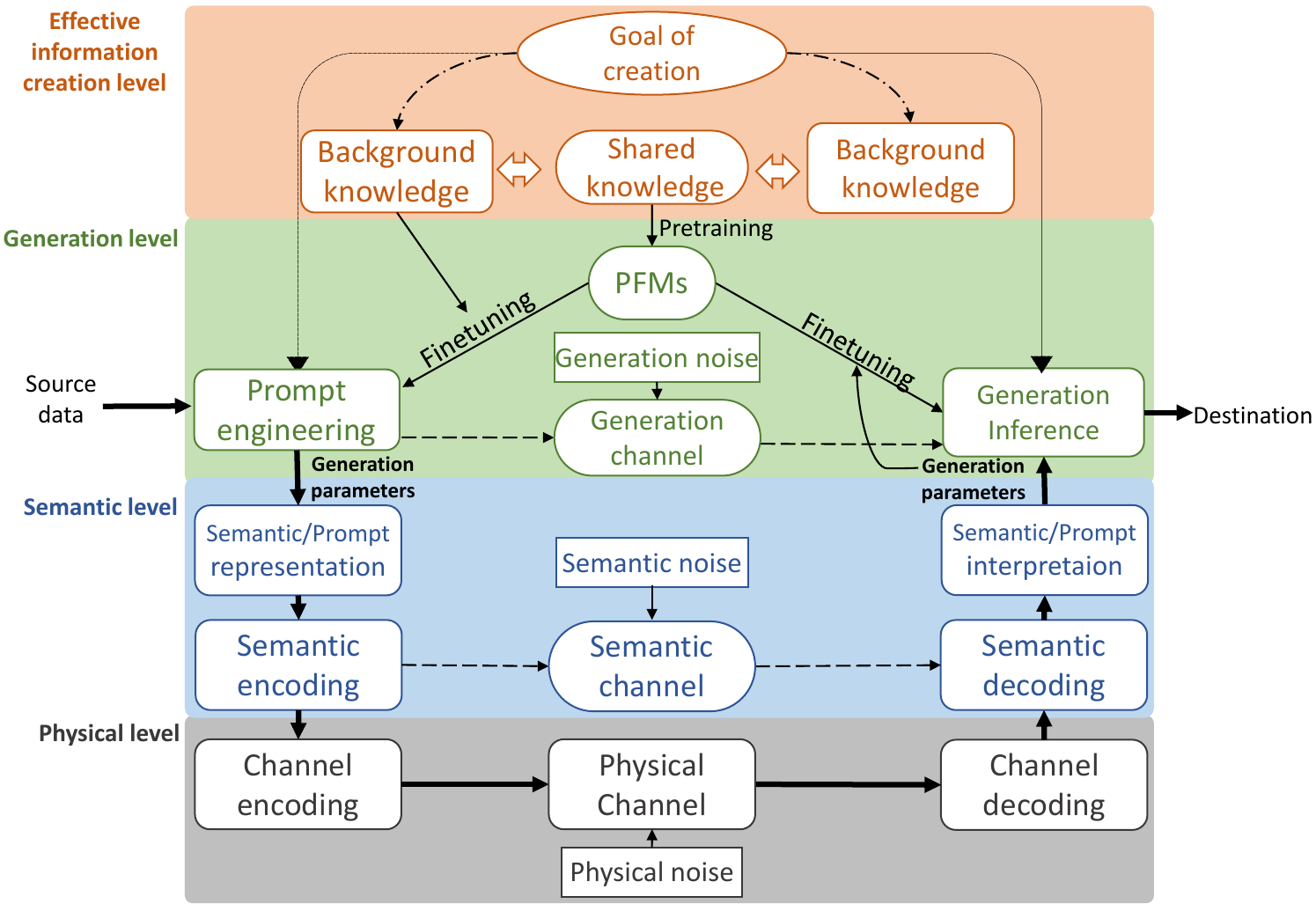}}
\caption{SemCom-enabled AIGC model, which is divided into physical level and semantic level for SemCom to support generation level and effective information creation level. }
\label{aigcsemcom}
\end{figure}
Traditional SemCom model prioritizes accurate transmission and reception of semantic content over relying solely on average information related to source data probabilities \cite{yangyang}. However, the advent of AIGC offers an opportunity to augment and benefit from SemCom model. In this section, we propose two integral components, the \textbf{{Effective Information Creation Level}} and the \textbf{{Generation Level}}. By implementing these components, AIGC can fulfill its dual objectives of information generation and communication more effectively.

\subsection{Integration of SemCom with AIGC}
In traditional SemCom model, as shown in Fig. \ref{aigcsemcom}, physical and semantic levels form the two-pronged structure.
{\textbf{The physical level}} encapsulates transmission operations within the physical layer, including modules responsible for channel encoding and decoding. This level necessitates symbol transmission over a physical channel susceptible to noise.
{\textbf{The semantic level}} is composed of semantic representation and encoding modules at the transmitter end. These components work together to extract and represent semantic information. Furthermore, semantic interpretation and decoding modules situated at the receiver end are for reconstructing and interpreting the received semantic information.

To extend the capabilities of the SemCom model, the introduction of {\textbf{{effective information creation level}}} and {\textbf{the generation level}} is essential \cite{yangyang}. Specifically, the {\textbf{effective information creation level}} in the AIGC framework transforms raw user data into semantic representations, which are then used to guide the generation process. It involves feature extraction and the creation of model fine-tuning parameters based on context-specific requirements. Meanwhile, the {\textbf{generation level}} provides an engine of content creation. It utilizes the semantically-encoded information and the generation parameters to run controlled diffusion models, generating the final content. Together, these levels augment the traditional SemCom model by adding control and personalization, thereby making AIGC a more adaptable and effective tool for content generation and transmission.

The generation level has the following components:
\begin{itemize}
\item \textbf{Background knowledge:} This component represents the foundation of knowledge and expertise necessary for content generation. On the transmitter side, it encompasses generation parameters that are finely tuned for specific generation tasks. On the receiver side, it includes the knowledge of how to fine-tune models with provided parameters and how to infer the generation process using the given generation parameters.
\item \textbf{Shared knowledge:} As a shared knowledge base, this component allows both the transmitter and receiver to access relevant information during content generation and interpretation. It consists of extensive data containing Pre-trained Foundation Models (PFMs) that can be utilized for additional fine-tuning, thereby enhancing the effectiveness and relevance of content generation \cite{zhou2023comprehensive}.
\item \textbf{Goal of creation:} This component represents the specific objective or intention behind content creation. It guides the generative process to ensure that the output aligns with the desired outcome.
\end{itemize}

The generation level has several integral components, including:
\begin{itemize}
\item \textbf{PFMs:} As the primary source of content generation, PFMs are pre-trained on extensive datasets \cite{zhou2023comprehensive}. This pre-training process endows them with considerable knowledge and capabilities, allowing them to be fine-tuned for specific tasks. Thus, they are capable of generating diverse, high-quality content such as images, stories and code.

Elaborating further on PFMs, it is important to understand that these models assimilate the underlying patterns from the vast amounts of training data. This wide-ranging and extensive training allows them to generate a rich variety of content, providing the basis for generating content in a variety of contexts.

\item \textbf{Prompt Engineering:} The process of creating prompts to guide AI model generation forms the crux of prompt engineering. By optimizing and customizing these prompts, the desired content output can be achieved. Prompt engineering plays a pivotal role in steering the generative process, resulting in more accurate and effective content generation.

The effectiveness of Prompt Engineering is determined by how well it aligns with the information generation task and the desired outcome. It necessitates a comprehensive understanding of the generative model and the task at hand to produce precise prompts that yield the most desirable output.

\item \textbf{Generation Inference:} This process involves content recreation at the receiver end, integrating and applying the generated content in the pertinent context.

\item \textbf{Generation Channel:} A virtual medium that facilitates content generation. This channel enables the transfer of creativity from the transmitter to the receiver.

\item \textbf{Generation Noise:} Representing potential inconsistencies or errors arising during the generative process \cite{zhou2023comprehensive}, this noise can lead to mismatches in the quality, values and utility of the generated content between the transmitter and receiver.
\end{itemize}






\subsection{Practical Applications and Insights}
The AIGC framework offers a significant advantage in terms of personalization, a feature that sets it apart from traditional SemCom. While traditional SemCom often remains constrained to transmitting generic semantic information, the SemCom-enabled AIGC framework exhibits the ability to tailor the generation of content to specific scenarios or user preferences such as personal driving assistant and smart healthcare \cite{du2023enabling}. 

To illustrate, consider a news organization employing the AIGC framework to generate digital content. Traditional SemCom model may merely transmit a generic scene of a car accident. However, the AIGC framework empowers the news organization (the receiver) to modify this scene according to the context of their report. The organization can integrate specific elements such as a particular street scene or additional prompts such as ``crowded street," thereby rendering the content more relevant and impactful for their audience, e.g., through personal driving assistant services. The framework thereby facilitates greater receiver control over the content, enhancing the value and applicability of the generative process.

Transitioning to data security considerations, the AIGC framework's feature of customization at the receiver's end becomes particularly advantageous. As transmitters only send semantic information and prompts, the receiver's specific adjustments remain undisclosed, significantly reducing the potential for inadvertent exposure of sensitive information. Thus, the AIGC framework proves especially suitable for applications requiring a high degree of privacy and data security, e.g., smart healthcare. Furthermore, the AIGC framework's potential to introduce generation noise actually contributes to its overall robustness. Although generation noise originates from inconsistencies within the AIGC model, service providers employing precise prompts and meticulous model fine-tuning techniques can mitigate these inconsistencies, yielding more reliable and high-quality generation results. Therefore, despite the inherent complexities of generation-based communication, the AIGC framework represents a robust mechanism, proficient in ensuring secure and personalized digital content generation and transmission.


\section{Controllable AI Generated Content services}
In this section, we explore the latest AIGC advancements, focusing on image generation from text prompts. We address how models such as ControlNet increases user control in the content generation process. Furthermore, we propose the application of controlled diffusion as generation encoder/decoder in the aforementioned generation framework, outlining the role of prompt engineering, service validation, receiver generation inference, and highlight how this shift improves data security and user experience.

\subsection{AIGC in Image Content Generation}
Our research primarily focuses on image content generation from text prompts in AIGC services, a recent development with transformative potential.
\begin{itemize}
\item \textbf{DALL·E:} Developed by OpenAI, DALL·E is a state-of-the-art AI model that demonstrates the capability to generate a wide array of images from text prompts \cite{pmlr-v139-ramesh21a}. This model represents a major step forward in translating abstract textual concepts into concrete visual imagery.

\item \textbf{VQGAN+CLIP:} In subsequent research, CLIP, a joint text-image encoder from OpenAI, was combined with a variant of Generative Adversarial Network (GAN) known as Vector Quantized GAN (VQGAN)\footnote{https://github.com/nerdyrodent/VQGAN-CLIP, accessed July 06, 2023}. This integration results in the capability of generating complex and creative images from text prompts.

\item \textbf{ControlNet+Stable diffusion:} Similar to DALL·E, Stable diffusion generates images from textual prompts. However, ControlNet enhances the user experience by providing more control over the generated output \cite{zhang2023adding}. By enabling modifications to the descriptive text during the generative process, ControlNet allows for more precise and interactive image creation.

\end{itemize}

These models, with their emphasis on high-resolution image synthesis, are revolutionizing the industry, lessening the reliance on individual skills and years of experience traditionally considered essential in this field. Therefore, they continue to progress and already significantly impact various domains, including advertising, game development and film effects. This transformative potential is paving the way for new standards and workflows.
\subsection{Enhanced Control in Generative AI }
One of the most significant challenges faced by generative models has been their limited controllability. Traditional methods have primarily relied on brute force techniques, involving linear combinations of prompts and the mass production of images. While this approach provides users with a broad scope of outputs, it also faces many issues, primarily rooted in inefficiency and a lack of precision.

However, the advent of ControlNet \cite{zhang2023adding} signals a transformative shift in AI-driven creation. ControlNet is adeptly integrated into the Stable Diffusion model, which is a large text-to-image diffusion model based on a U-Net architecture. By using a unique dual-parameter mechanism involving a locked and trainable copy of parameters, ControlNet ensures efficient learning while preventing overfitting. Moreover, it utilizes zero convolution layers as a key technique, providing an efficient connection between different network blocks. Even though these layers start with zero weights, they adapt during the training phase, contributing to the network's output and enhancing control. This structure allows for layer-wise manipulation of the Stable Diffusion model, adding controllability to the image generation process.

ControlNet allow users to interactively guide the image generation process. By adding control over the neural network's behavior by manipulating the input conditions of network blocks, ControlNet provides another modality of control in AI-based content creation. In essence, ControlNet improved control and precision capabilities. For example, a generated dancing scene can be controlled with the dedicated pose extracted from a guiding image. 

\subsection{Controlled Diffusion as the generation Encoder/Decoder}\label{faef}
Interestingly, the shift towards controlled generation aligns closely with the core principles of the proposed framework. SemCom inherently involves the extraction and construction of information, paralleling the controlled generative process as exemplified by controlled diffusion technologies. This characteristic alignment underscores the potential synergy between generative AI models and SemCom, indicating the vast opportunities present for their integration. Given the idea, The data flow across the proposed information generation and communication can be described through the following steps:
\begin{figure*}[t!]
\centerline{\includegraphics[width=1\textwidth]{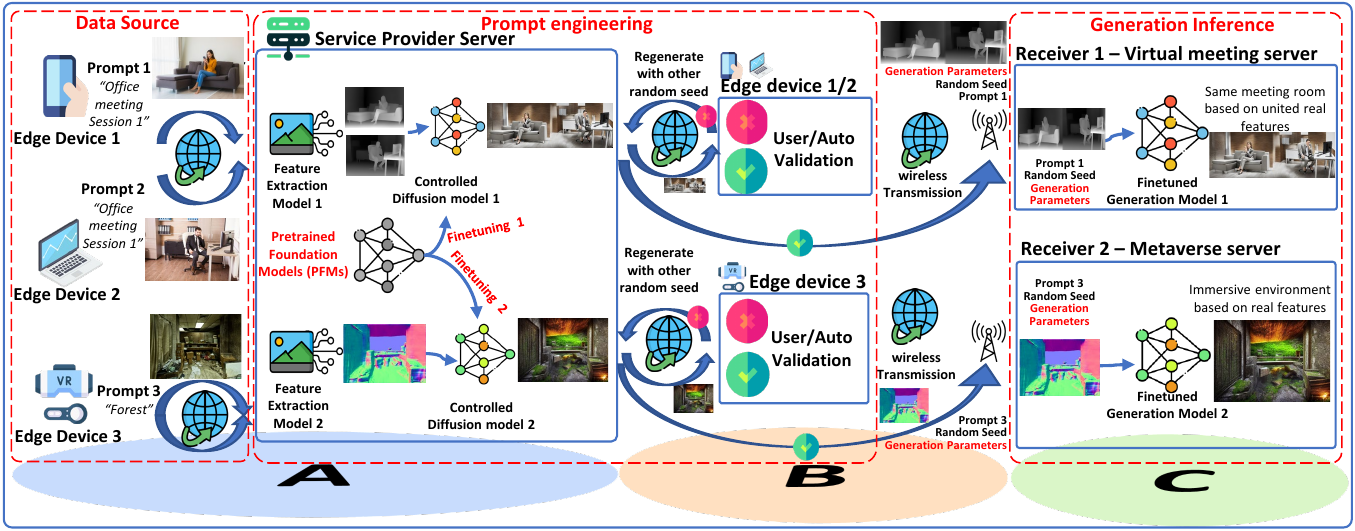}}
\caption{AIGC model as generation encoder and decoder. In this AIGC framework, edge devices 1 and 2 represent users in a professional meeting. This step fuses their \textbf{Semantic/prompt representation} through \textbf{Prompt engineering}, guiding AI models to generate a unified modern meeting room theme in \textbf{Generation inference}. On the other hand, edge device 3 applied this framework in VR to generate immersive environments and to synchronize with metaverse server, addressing VR bounding issues by automatically adapting to the user's surroundings environments.}

\label{iot}
\end{figure*}

\textbf{Step 1. Prompt engineering :} As illustrated in Fig. \ref{iot} Part A, the transmission process commences with data collection by the edge devices, which can encompass various user interfaces such as sensors and smartphones. These devices gather raw user inputs, ranging from images and text prompts to user-specific data like the service type that ultimately defines the generation prompt, and transmit this data to a server. This server is a specialized unit designed for feature extraction which transforms the raw data into an abstract semantic representation. The server then employs controlled diffusion models guided by a collection of generation parameters including the random seed and generation steps, among other settings related to the generative process. Subsequently, these models will generate content derived from the given prompts and extracted semantic information. Within this framework, Pretrained Foundation Models (PFMs) play a critical role as a shared knowledge base, accessible to both the transmitter and receiver, thus instead of transmitting entire large pretrained models, only the transmission of the fine-tuning parameters is required. This strategic approach significantly reduces communication overhead and enhances the overall creation efficiency. Furthermore, the procedure is designed to be shareable across multiple edge devices, an attribute that optimizes resource allocation and maximizes system performance. The outcome of this stage is a semantic/prompt representation—an encoded version of the user's input that encapsulates both the extracted features and the generation parameters.

\textbf{Step 2. Servcie validation:} As shown in Figure \ref{iot} Part B, upon generating the content, the server conveys it back to the transmitter for validation. This step can either be supervised by the user or automated mechanism through predefined criteria or algorithms such as convolutional neural networks for image recognition. Though it may create more communication overheads, this validation process is integral to the framework, as it not only enhances communication but also enables creative possibilities. The validation process fortifies the reliability and quality of the service, ensuring that the produced images meet the established standards and user requirements. If validation criteria is met, the corresponding semantic/prompt representation is then dispatched for transmission. Conversely, if the content does not meet validation standards, regeneration is initiated with adjustments made to the generation variables. Traditional SemCom model views Parts A and B together as a semantic encoder, transforming user inputs into semantically encoded information. However, within the context of the AIGC framework, this process is interpreted as prompt engineering. This process not only includes the encoding of semantic content but also encompasses the formulation and refinement of generation prompts, thus ensuring a more precise, efficient and user-specific content generation. In addition, the validation process could be adopted in determining the minimal prompt/semantic representation necessary for recreating content.  Aligned with Semantic Entropy(SE) \cite{10001594}, such a concept can be termed as ``generation entropy". This concept concurs with the process of prompt engineering and ensures maximum overhead reduction during transmission.

\textbf{Step 3. Receiver generation inference:} Following successful validation, the semantic/prompt representation is sent to the receiving entity, which can be another edge device or a server, as depicted in Fig. \ref{iot} Part C. In contrast to the traditional SemCom model, where this step serves as a semantic decoder, the AIGC framework interprets it as an embodiment of generation inference, thus underlining the creative aspect of content interpretation from the received information. The versatility of this process is demonstrated across various domains. For example, in gaming, it facilitates the creation of user-specific avatars using the received semantic information. In the realm of virtual reality, this technology addresses the common issue of obstructed vision within VR bounding areas\footnote{https://github.com/Zetaphor/webxr-environment-mapper, accessed July 28, 2023}. Rather than disrupting the user experience with the necessity of manually drawing a bounding box, the system can regenerate an environment reflecting the user's actual surroundings, thereby enhancing the immersive experience.

In conventional content generation methods, transmission of sensitive user data, such as images or videos, over the Internet potentially risks user privacy. However, the AIGC model operates as an encoder-decoder system for SemCom, transmitting only semantic information and generation variables. This method maintains the transmitted data encrypted and inaccessible to unauthorized entities equipped with different AIGC models and generation parameters, thereby amplifying data privacy and security. For example, eavesdroppers may not have the correct AIGC inference engine, unable to generate the same or similar content as the legitimate receiver. Moreover, this approach transitions power to users, offering a versatile tool in place of a predetermined outcome. 

\begin{figure*}[t!]
\centerline{\includegraphics[width=1\textwidth]{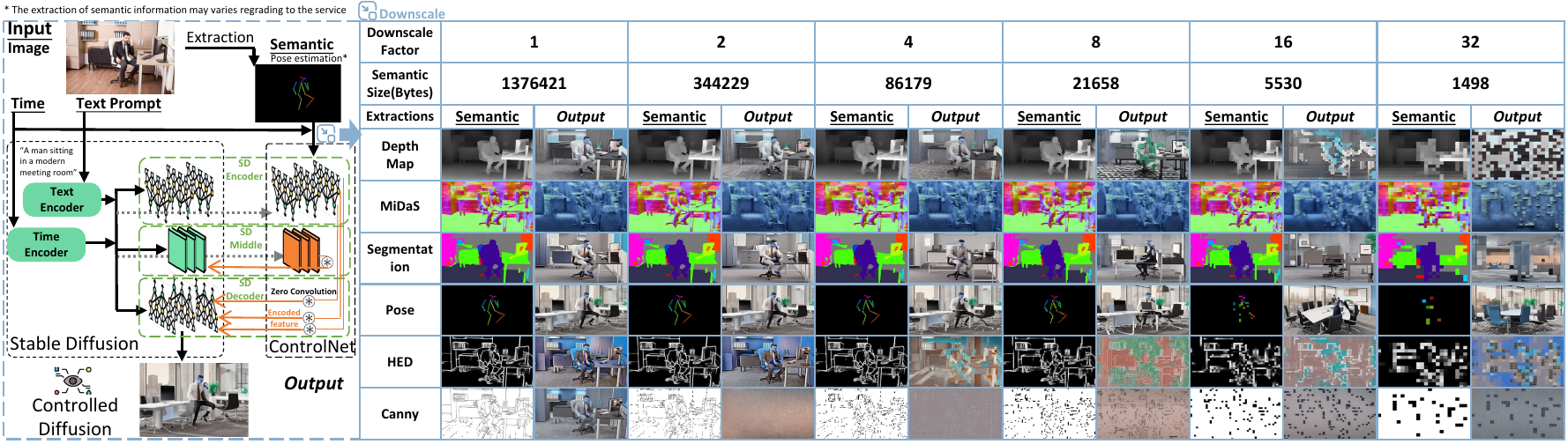}}
\caption{Semantic variation and downscaling impact in generative task. The figure illustrates ControlNet's structure and where downscaling in semantic information could be deployed. Different extraction models show varied scaling levels, reducing transmitted data. Each method exhibits different sensitivity to downscaling, highlighting the need for flexible resource allocation. Tailoring the downscale factor to the specific model ensures efficient communication by prioritizing less sensitive to more sensitive methods. }
\label{systemdigram}
\end{figure*}
\section{Case Study: Joint Optimization of SemCom and AIGC}
This section presents a case study on the joint optimization of SemCom and AIGC services based on ControlNet\cite{zhang2023adding}. We investigate most suitable semantic extraction methods, the role of evaluation metrics in content quality maintenance and the challenges of joint resource allocation across multiple components, from physical, semantic and generation levels. Through empirical studies and the use of DQN, this section offers insights and solutions for more efficient resource allocation in the multifarious realm of AIGC services.
\subsection{Semantic extraction methods}

Semantic extraction is a crucial component of the transmission framework of AIGC services. The method of semantic extraction depends primarily on the type of service and can vary considerably. As depicted in Fig. \ref{systemdigram}, several common semantic information such as depth maps, Multi-Instance Depth via Attention Sampling (MiDaS), segmentation, pose estimation, Holistically-Nested Edge Detection (HED) and Canny edge detection have been employed in our proposed transmission framework. Depth maps typically measure the distance between the imaging sensor and each pixel's corresponding real-world point, resulting in a grayscale image representation. Meanwhile, MIDAS uses a novel attention mechanism to produce high-quality depth maps from 2D images \cite{DBLP:journals/corr/abs-2111-00947}. On the other hand, segmentation separates an image into regions or objects for further analysis. For pose estimation, which interprets the human form in an image or video, recognizing human posture and outputting skeleton position information. Moreover, HED leverages the power of deep learning to predict the presence and location of edges in an image, providing a structural layout. In contrast, Canny edge detection is a multi-stage algorithm used to detect a wide range of image edges, offering a binary output that emphasizes the boundaries of objects within an image \cite{DBLP:journals/corr/XieT15}. These examples represent a fraction of the broad, flexible range of methods to obtain semantic representation (As discussed in Section~\ref{sec2}) that services can deploy. 

A case in point is when a service pertains to human figures, where pose estimation semantic information have the potential to retain substantial information. This allows for satisfactory reconstruction results at the receiver's end. In our experimental studies, we found that the final output only diverged significantly from the original text prompt, ``a man sitting in a modern meeting room" when the semantic information is subjected to a downscaling factor of 10. This observation suggests that even under significant compression, the chosen semantic (pose) preserved essential information effectively.

However, the selection of semantic information may only sometimes align with the service type or user requirements. For instance, another scenario outlined in Fig. \ref{systemdigram} used a depth map as the semantic information for a different service. In this case, the maximum permissible downscaling factor without noticeable loss of information is only 4, corresponding to image compression of 16 times. Therefore, the degree of compression semantic information can be adapted without impacting quality depends heavily on the chosen semantic information. This highlights the importance of carefully selecting the most suitable semantic extraction method based on service type and user requirements.

\begin{figure*}[t]
\centerline{\includegraphics[width= 1\textwidth]{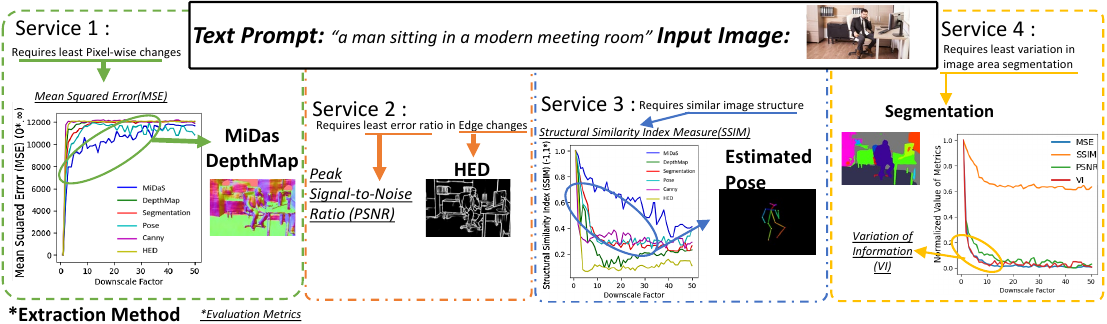}}
\caption{Workflow for defining predictable pairs of semantic extraction methods and evaluation metrics. Service 2 presents a traditional case where a service provider defines both the extraction method and evaluation metric. Services 1 and 3 illustrate the process of selecting the most suitable semantic extraction method when the service provider defines the evaluation metrics. Conversely, Service 4 demonstrates the process of choosing the most appropriate evaluation metrics when the semantic extraction method is defined. This iterative workflow ensures the identification of most predictable combinations for effective semantic extraction and evaluation in the given context.}
\label{ssim}
\end{figure*}
\subsection{Evaluation metrics}
The proposed methodology for identifying the most suitable semantic extraction method and corresponding evaluation metrics for a particular service is illustrated in Fig. \ref{ssim}, using four service examples. 

For Services 1 and 3, the process begins by selecting an evaluation metric aligning with the provider's quality assessment needs. This step is followed by identifying a semantic extraction method that pairs effectively with the chosen metric, particularly the semantic extraction method and evaluation metric pair that is providing a predictable and stable linear response to various levels of image degradation. This approach ensures predictable and maintainable content quality, even under diverse degrees of downscaling. On the contrary, Service 4 employs a reversed process where the service provider selects the most suitable semantic extraction method that aligns with the service type or user requirements and subsequently identifies an evaluation metric offering a predictable linear response to the chosen extraction method.

Service 2 presents a different scenario, where the service provider simultaneously defines both the extraction method and evaluation metric. This approach is advantageous when service requirements are well-understood and quality standards are clearly established.

This study utilizes a variety of established evaluation metrics, including Mean Squared Error (MSE), Peak Signal to Noise Ratio (PSNR), Structural Similarity Index (SSIM) and Variation of Information (VI) \cite{5596999}. MSE measures the average pixel-wise squared difference between the generated images with and without semantic downscaling. PSNR compares the maximum possible power of the original signal to the power of the noise that affects its quality. SSIM assesses structural similarity which provides insight into perceived changes in structural information and VI quantifies the change in shared information, reflecting the effectiveness of semantic content transmission. Although other metrics, such as Inception Score (IS), Fréchet Inception Distance (FID) \cite{benny2021evaluation} and Content Loss \cite{chen2023learning} could be considered, this study concentrates on a selected group of common metrics for simplicity and clarity.

In summary, selecting the most predictable pair of semantic extraction methods and evaluation metrics can be complex and service-specific. However, a systematic approach that considers the provider's needs and the pair's linear response under compression can yield predictable image quality.

\subsection{Multi service joint resource allocation}

Following the selection of distinct semantic extraction methods and evaluation metrics tailored to individual service , a complex dynamic emerges when accommodating multiple services' transmission requirements concurrently. To provide a comprehensive understanding of our framework's deployable environment, we detail the specifications of the equipment specification used in our experiments:
\begin{itemize}
  \item CPU: AMD Ryzen Threadripper PRO 3975WX 32-Cores
  \item GPU: NVIDIA RTX A5000
\end{itemize}
\begin{figure}[h]
  \centering
  \subfigure[Reward analysis]{\includegraphics[width=0.2\textwidth]{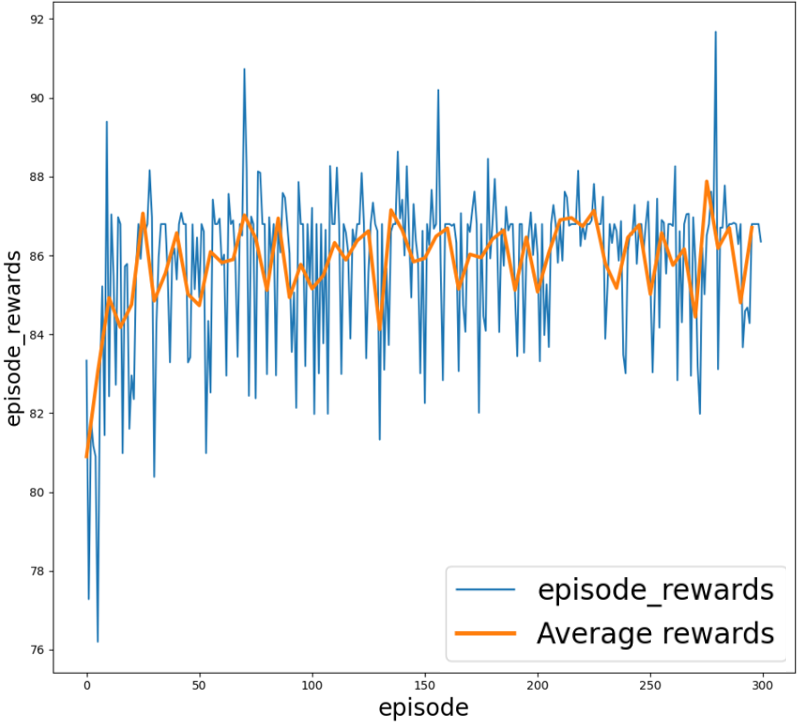}    \label{fig:dqnREWARD}}\quad
  \subfigure[Loss analysis]{\includegraphics[width=0.2\textwidth]{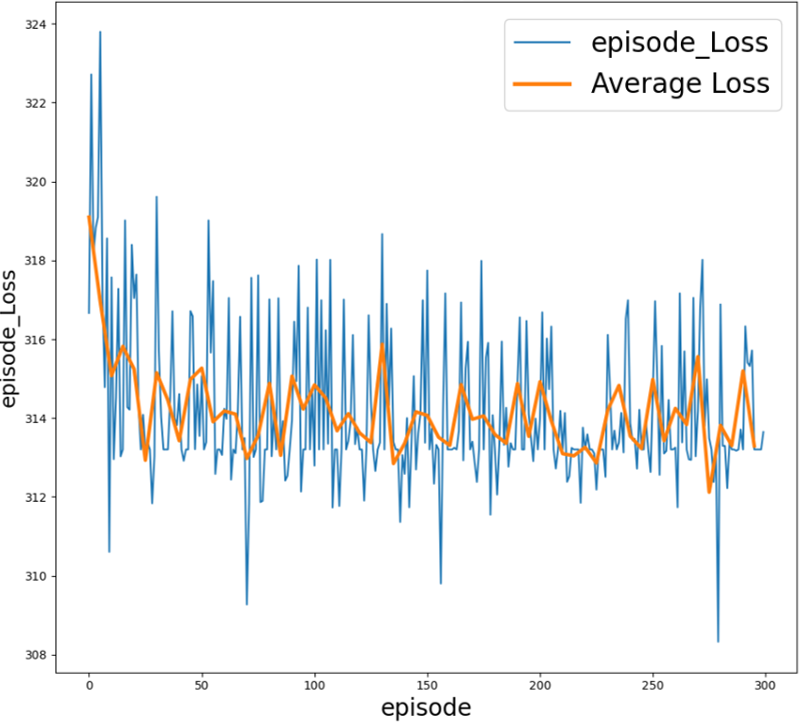}\label{fig:dqnLOSS}}
   \caption{Rewards and down-sampling losses over episodes.}
  \label{fig:rewardandloss}
\end{figure}

The scenario, as represented in Fig. \ref{ssim}, involves four services, each possessing unique evaluation metrics and semantic extraction methods. For the purpose of maintaining consistency across varying metrics, all evaluation metrics were normalized to fall within a range of zero to one. Here, a value of one denotes an exact match between the original and regenerated image, indicating no loss of information or quality. In this study, we employed Deep Q-Learning Network (DQN) as a solution technique to assess the feasibility of joint resource allocation. In the DQN model, the action space is defined by the downscaling factor for each service, while the reward signal is measured based on the evaluation results.

The outcomes of the DQN optimization are depicted in Fig. \ref{fig:rewardandloss}. As the results indicate, the optimization process converges after approximately 50 epochs. This convergence is significant as it indicates the attainment of the most efficient resource allocation scheme, which is essential for maximizing the efficiency and effectiveness of our AIGC framework. The loss is calculated by comparing the quality of images generated with original and down-scaled semantics, reflecting how resource allocation schemes affects image quality. To enhance system performance, the reward is inversely modeled related to loss, guiding the DQN to efficiently allocate resources while preserving image quality by reducing loss. The substantial fluctuation observed within this process is a likely consequence of the escalated complexity of image restoration when the downscaling factor increments. Specifically, an increased downscaling factor signifies a greater division of the image's width and height. This substantial partitioning invariably complicates restoring the original semantic information, thereby influencing the quality of the final generated image. 

Nonetheless, despite these challenges, the results validate the feasibility of the proposed framework for joint resource allocation. The approach demonstrates adaptability to the diverse range of requirements inherent in contemporary and future AIGC services. Therefore, this framework holds significant potential for widespread deployment, facilitating more efficient and optimized resource allocation within the diverse landscape of AIGC services.

In Section I, we highlighted AIGC + SemCom's capability to manage complex content generation, which is later substantiated by our experiments in Section IV-B. Our model adeptly managed resource allocation for four distinct services, showcasing its effectiveness in complex scenarios that align with our theoretical assertions about the AIGC + SemCom framework's practicality in real-world applications. Additionally, using the DQN is pivotal in determining the most suitable allocation schemes. This strategy efficiently downscales semantics for transmission with the least effect on generated image quality, demonstrating efficient and high-quality content creation. Such results affirm the framework's efficiency and quality standards.
{



}
\section{Future Research Directions}
In this section, we present future research directions related to the integration of SemCom and AIGC services:
\begin{itemize}
  \item \textbf{Development of Universal Evaluation Metrics}: Current AIGC services present a challenge in creating universal evaluation metrics, as different services have varying requirements~\cite{du2023enabling}. A comprehensive metric that quantifies performance across various AIGC services can be developed to provide a consistent performance measure independent of the services. This would enable easier comparison across different services and provide a better understanding of performance trade-offs in various AIGC applications.

  \item \textbf{Exploration of Low-rank Adaptation (LoRa)}: Investigate the application of LoRa, a method that condenses large language models into smaller, efficient versions without significant performance loss~\cite{hu2021lora}. By leveraging LoRa's capabilities to optimize image creation and fine-tuning, it is possible to enhance the transmission and decoding of semantic information in AIGC services. This approach can improve the accessibility and usability of AIGC services in resource-limited environments by providing streamlined and efficient fine-tuning of PFMs, making the process more standardized and effective.

  \item \textbf{Development of Lightweight AIGC Services}: 
  Adapting AIGC services to be more lightweight and efficient for mobile edge computing can be a potential direction. This direction focuses on optimizing AIGC algorithms to ensure they are resource-efficient and can operate effectively in the constrained environments typical of edge devices for a better quality of the service~\cite{verma2019energy}. Such a method aims to enhance real-time data processing and content generation capabilities in edge computing scenarios, which could contribute to advancements in smart transportation and connected vehicle technologies.
\end{itemize}

\section{Conclusion}
This article has presented a conceptual model for integrating SemCom into AIGC services, featuring an additional generation level for content creation. Our analysis has revealed how modifications in input information extraction, like image resolution and compression, can impact AIGC output. The introduced evaluation framework highlighted AIGC’s role in semantic information transmission and facilitated optimized resource allocation. Validation through a Deep Q-Network confirmed the feasibility of our framework, promising significant efficiency improvements for AIGC services.

\bibliographystyle{IEEEtran}
\bibliography{main}
\end{document}